\documentclass[11pt]{article}
\usepackage[dvips]{graphicx}
\usepackage{psfrag}
\usepackage{amssymb}
\usepackage{amsmath}
\usepackage{amscd}
\usepackage{latexsym}
\usepackage{bbm}

\catcode`@=11
\renewcommand{\section}{\@startsection{section}{1}{0pt}{\medskipamount}
{\medskipamount}{\large\bf}}
\numberwithin{equation}{section}
\catcode`@=12
\def\a{\alpha}
\def\b{\beta}
\def\g{\gamma}

\def\h{\eta}
\def\th{\theta}

\def\la{\lambda}

\def\s{\sigma}
\def\p{\phi}
\def\vp{\varphi}

\newcommand{\unity}{\mathbbm{1}}

\def\im{\textrm{i}}
\def\ep{\textrm{e}}
\def\diff{\textrm{d}}

\def\pa{\partial}
\def\rd#1{\buildrel{_{_{\hskip 0.01in}\rightarrow}}\over{#1}}
\def\ld#1{\buildrel{_{_{\hskip 0.01in}\leftarrow}}\over{#1}}
\def\sfrac#1#2{{\textstyle\frac{#1}{#2}}}
\def\>{\rangle}
\def\<{\langle}
\def\+{\dagger}
\def\={\ =\ }
\def\und{\qquad\textrm{and}\qquad}
\def\tU{\textrm{U}}
\def\vel{\textrm{v}}
\def\Tdag{T^{\dagger}}
\def\gb{\bar{\gamma}}
\def\tN{\widetilde{N}}
\def\tP{\widetilde{P}}

\begin{document}

\begin{titlepage}
\setcounter{page}{0}
\begin{flushright}
ITP--UH--18/08\\
\end{flushright}

\vskip 2.0cm

\begin{center}

{\Large\bf  The noncommutative sine-Gordon breather
}

\vspace{12mm}

{\large Andr\'e Fischer \ and \ Olaf Lechtenfeld
}
\\[8mm]
\noindent {\em Institut f\"ur Theoretische Physik,
Leibniz Universit\"at Hannover \\
Appelstra\ss{}e 2, 30167 Hannover, Germany }\\
{Emails: afischer, lechtenf @itp.uni-hannover.de}

\vspace{12mm}

\begin{abstract}
\noindent
As shown in [hep-th/0406065], there exists a noncommutative deformation 
of the sine-Gordon model which remains (classically) integrable
but features a second scalar field. We employ the dressing method
(adapted to the Moyal-deformed situation) for constructing the deformed
kink-antikink and breather configurations. Explicit results and plots
are presented for the leading noncommutativity correction to the breather.
Its temporal periodicity is unchanged.
\end{abstract}

\end{center}
\end{titlepage}

\section{Introduction and summary}

\noindent
The sine-Gordon model is a paradigm for relativistic integrable models in 
$1{+}1$ dimensions (e.g., see~\cite{raja}.) Its multi-soliton spectrum
is well known and consists not only of multi-kink scattering configurations
but also of bound states, the simplest of which is the so-called breather.
It may be obtained formally by analytically continuing the kink-antikink
configuration in its relative velocity variable, $\vel\to\im\vel$,
and oscillates periodically in time.

A systematic procedure for deriving the integrability features of the 
sine-Gordon model relates it to the self-duality equations of SU(2) Yang-Mills
theory in $2{+}2$~dimensions~\cite{dimred}. In a light-cone gauge, 
these equations follow from the Nair-Schiff action~\cite{nair}. 
A first and straightforward dimensional reduction produces Ward's modified 
chiral sigma-model action for SU(2)-valued fields 
in $2{+}1$ dimensions~\cite{ward}.
A second dimensional reduction then generates an abelian sigma model
equivalent to the sine-Gordon theory. To arrive there, one must prescribe
a particular dependence on one spatial coordinate (rather than trivial
independence) and also algebraically restrict the field from SU(2) to a
U(1)~subgroup. The remaining phase~$\vp(t,x)$ turns out to be ruled by
the sine-Gordon equation, with the coupling or mass~$\alpha$ appearing 
as a parameter of the dimensional reduction.

For several years now the Moyal deformation of integrable field theories
has been of some interest. 
In particular, the Ward model~\cite{lepo01,goteborg,chu} and the sine-Gordon
model~\cite{lmppt,kule} have been generalized to the noncommutative realm.
The key insight for the latter case was that the extension of SU(2) to~U(2),
necessary for implementing the Moyal deformation in the Yang-Mills theory,
should be retained under the dimensional and algebraic reduction, so that
the noncommutative sigma-model field takes its values in U(1)${\times}$U(1)
rather than U(1). The deformed sine-Gordon model so obtained~\cite{lmppt}
features {\sl two\/} scalar fields (phases) $\p_+$ and~$\p_-$, whose 
noncommutative abelian WZW~actions are coupled in a simple way. 
In the commutative limit, the average $\vp=\sfrac12(\p_+{+}\p_-)$ of these 
phases produces the standard sine-Gordon field while their difference
$\p_+{-}\p_-$ decouples as a free field.

Since the powerful techniques for constructing multi-soliton solutions
in integrable models have been shown to survive the noncommutative
deformation, it is straightforward (but may be tedious) to work out
such configurations for the Moyal-deformed sine-Gordon model as well.
The basic strategy was already outlined in~\cite{lepo01}
but has been applied only to the simple kink so far~\cite{lmppt}. 
However, owing to the relativistic invariance, a one-kink configuration
depends only on its single co-moving coordinate~$\h(t,x)$, and so it cannot 
get deformed. Only multi-lumps with relative motion should be affected by 
noncommutativity. The first instances are the two-kink, kink-antikink and
breather solutions.

In this letter we apply the Moyal deformation to the two latter cases.
It is important to verify the effect of noncommutativity, since the tree-level
computations of~\cite{lmppt} had suggested that perhaps the entire Moyal 
deformation of the sine-Gordon model might be ficticious. Here, we demonstrate
this {\sl not\/} to be the case, by working out the first-order 
(in the noncommutativity parameter) correction to the `classical'
kink-antikink and breather configurations. It turns out that this leading
correction affects only the would-be free field $\p_+{-}\p_-$;
the generalized sine-Gordon field~$\vp$ gets modified at second order onwards,
as does the energy density.
Only the substantial calculational effort prevented us from evaluating 
higher orders, but we present the starting-point equations for doing so.
As an exact result, the temporal periodicity of the breather is unchanged
by the deformation.

\bigskip

\section{The model}

\noindent
The integrable noncommutative sine-Gordon model introduced in~\cite{lmppt}
involves {\it two\/} U(1)-valued fields
\begin{equation}
g_+(t,x)\=\ep_\star^{ \frac{\im}{2}\p_+(t,x)}\ \in \tU(1)_+ \und 
g_-(t,x)\=\ep_\star^{-\frac{\im}{2}\p_-(t,x)}\ \in \tU(1)_-
\end{equation}
and may be defined via its action
\begin{equation} \label{gaction}
S[g_+,g_-] \= S_{\textrm{wzw}}[g_+]\ +\ S_{\textrm{wzw}}[g_-]\ +\ 
\alpha^2\!\int\!\diff{t}\,\diff{x}\;\bigl(g_+^\+g_-+g_-^\+g_+-2\bigr) \quad,
\end{equation}
where $S_{\textrm{wzw}}$ is the abelian WZW action
\begin{equation} \label{WZWaction}
S_{\textrm{wzw}}[g]\= - \sfrac12\int\!\diff t\,\diff x\; 
\bigl( \pa_t g^\+\,\pa_t g - \pa_x g^\+\,\pa_x g) \ -\ 
\int\!\diff t\,\diff x\int_0^1\!\!\diff \la\ 
\hat g^\+\pa_{[t}\hat g\star\hat g^\+\pa_{x]}\hat g\star\hat g^\+\pa_\la\hat g
\end{equation}
with a homotopy path $\hat{g}(\la)$ connecting
$\hat{g}(0)=1$ and $\hat{g}(1)=g$ and a Moyal star product
\begin{equation}
(f_1 \star f_2)(t,x)\ =\ f_1(t,x)\,\exp\,\bigl\{ \sfrac{\im}{2}
({\ld{\partial}}_t \,\theta\, {\rd{\partial}}_x -
 {\ld{\partial}}_x \,\theta\, {\rd{\partial}}_t \bigr\}\,f_2(t,x) 
\qquad\textrm{so that}\quad
[t,x]_\star = \im\,\theta \quad.
\end{equation}

In light-cone variables
\begin{equation} \label{lightcone}
u\ :=\ \sfrac{1}{2}(t+x)\quad,\qquad 
v\ :=\ \sfrac{1}{2}(t-x)\quad,\qquad
\pa_u\ =\ \pa_t+\pa_x\quad,\qquad
\pa_v\ =\ \pa_t-\pa_x
\end{equation}
the corresponding equations of motion read
\begin{equation} \label{geom}
\begin{aligned}
\pa_v\bigl(g_+^\+\star\pa_u g_++g_-^\+\star\pa_u g_-\bigr) &\= 0 \quad,\\[6pt]
\pa_v\bigl(g_+^\+\star\pa_u g_+-g_-^\+\star\pa_u g_-\bigr) &\=
2\a^2\bigl(g_+^\+\star g_- - g_-^\+\star g_+ \bigr) \quad,
\end{aligned}
\end{equation}
which in the commutative limit $\theta{\to}0$ simplifies to
\begin{equation} \label{ceom}
\pa_u\pa_v (\p_+{-}\p_-) \= 0 \qquad\textrm{and}\qquad
\pa_u\pa_v (\p_+{+}\p_-) \= -8\a^2\,\sin\sfrac12(\p_+{+}\p_-) \quad.
\end{equation}
Hence, the identification of the standard sine-Gordon field~$\vp$ 
with mass $2\a$ is made via
\begin{equation}
\sfrac12(\p_+{+}\p_-) \= \vp \ +\ O(\theta) 
\qquad\textrm{or}\qquad
g_-^\+\star g_+ \= \ep^{\im\vp} \ +\ O(\theta) \quad.
\end{equation}
For later use, we embed the U(1)-valued fields into U(2),
\begin{equation} \label{embed}
G\ :=\ \sfrac12 \, \biggl(\begin{matrix}
g_+ {+} g_- & \ \ g_+ {-} g_- \\[4pt]
g_+ {-} g_- & \ \ g_+ {+} g_- 
\end{matrix}\biggr) \quad\buildrel{\theta\to0}\over{\longrightarrow}\quad 
\ep^{\frac{\im}{4}(\p_+-\p_-)} \biggl(\begin{matrix}
\  \cos\sfrac{\vp}{2} & \im\sin\sfrac{\vp}{2} \\[4pt] 
\im\sin\sfrac{\vp}{2} & \  \cos\sfrac{\vp}{2}
\end{matrix}\biggr) \quad.
\end{equation}

\bigskip

\section{Dressing construction}

\noindent
The breather solution may be obtained from a kink-antikink configuration
with relative velocity $2\vel$ by an analytic continuation $\vel\to\im\vel$.
The co-moving coordinates for the kink and antikink read
\begin{equation}
\h_1 \= +p\,u-\sfrac1p\,v \= +\g\,(x-\vel t) \und
\h_2 \= -\sfrac1p\,u+p\,v \= -\g\,(x+\vel t) \quad,
\end{equation}
respectively, where $p\in(0,1)$, 
\begin{equation}
\vel \= \sfrac{1-p^2}{1+p^2} \ >0 \qquad\Leftrightarrow\qquad
p^2 \= \sfrac{1-\vel}{1+\vel} \und
\g \= \sfrac{1}{\sqrt{1-\vel^2}} \= \sfrac12\bigl(p+\sfrac1p\bigr) \quad.
\end{equation}

A convenient way to construct the kink-antikink solution employs the
dressing method. For the case at hand, it yields~\cite{lepo01}
\begin{align}
G \= \unity\ &-\ 
2\,\bigl( \unity + \sfrac{1-\vel}{\vel}\,P_2 \bigr) \star T_1\star
\bigl[\Tdag_1\star(\unity{-}\s P_2)\star T_1\bigr]_\star^{-1}\star\Tdag_1
\nonumber \\[6pt] & -\ 
2\,\bigl( \unity - \sfrac{1+\vel}{\vel}\,P_1 \bigr) \star T_2\star
\bigl[\Tdag_2\star(\unity{-}\s P_1)\star T_2\bigr]_\star^{-1}\star\Tdag_2
\label{Gconf} \\[12pt] =\ \unity\ &-\ 
2\,\bigl( \unity + \sfrac{1-\vel}{\vel}\,P_2 \bigr) \star 
P_1 \star \bigl[ \unity{-}\s P_2\star P_1\bigr]_\star^{-1} -\
2\,\bigl( \unity - \sfrac{1+\vel}{\vel}\,P_1 \bigr) \star
P_2 \star \bigl[ \unity{-}\s P_1\star P_2\bigr]_\star^{-1}
\quad, \nonumber
\end{align}
where $\s=\frac{-4p^2}{(1-p^2)^2}=1-\vel^{-2}$ 
and we introduced hermitian projectors
\begin{equation} \label{proj}
P_1 \= T_1\star(\Tdag_1\star T_1)_\star^{-1}\star\Tdag_1 \und
P_2 \= T_2\star(\Tdag_2\star T_2)_\star^{-1}\star\Tdag_2
\end{equation}
based on $2{\times}1$ matrix-valued functions $T_1(\h_1)$ and $T_2(\h_2)$
related to the kink and antikink components of the configuration.
The $T_i$ are determined only up to right (star-) multiplication with an
arbitrary invertible function and may be taken as
\begin{equation}
T_1 \= \biggl(\begin{matrix} 1 \\[4pt]  \im\ep^{2\a\h_1} \end{matrix}\biggr)
\und
T_2 \= \biggl(\begin{matrix} 1 \\[4pt] -\im\ep^{2\a\h_2} \end{matrix}\biggr)
\end{equation}
by a suitable choice of the coordinate origin.
Note that we have dropped the star index on the exponentials since each one
depends on a single coordinate combination only.

By inserting these $T_i$ into (\ref{proj}), the ensuing projectors into
(\ref{Gconf}) one is in principle able to read off $g_\pm$ from~(\ref{embed}) 
and extract the noncommutative breather configuration $\p_\pm$.

\bigskip

\section{Commutative breather}

\noindent
Before delving into the explicit computation, let us first retrieve the
familiar commutative breather in the $\theta{\to}0$ limit.

Since a coordinate rescaling modifies the coupling~$\a$ we take the freedom
to put $2\a=1$ in the following. Dropping all stars, one first builds
\begin{equation} \label{projconf}
P_1 \= \frac{1}{1+\ep^{2\h_1}}\ \biggl(\begin{matrix}
1 & -\im\ep^{\h_1} \\[4pt] \im\ep^{\h_1} & \ep^{2\h_1} 
\end{matrix}\biggr) \und 
P_2 \= \frac{1}{1+\ep^{2\h_2}}\ \biggl(\begin{matrix}
1 & \im\ep^{\h_2} \\[4pt] -\im\ep^{\h_2} & \ep^{2\h_2}
\end{matrix}\biggr) 
\end{equation}
and thus
\begin{equation}
\bigl[\Tdag_j\,(1{-}\s P_k)\, T_j\bigr]^{-1} \=
\frac{1+\ep^{2\h_k}}
{(1{-}\s)(1{-}\ep^{\h_1+\h_2})^2+(\ep^{\h_1}{+}\ep^{\h_2})^2} 
\qquad\textrm{for}\quad (j,k) = (1,2)\ \textrm{or}\ (2,1) \quad.
\end{equation}
Next, one obtains 
\begin{equation}
\begin{aligned}
\bigl(\unity+\sfrac{1-\vel}{\vel}\,P_2 \bigr)\,T_1\,\Tdag_1 \=&\
\frac{1}{1+\ep^{2\h_2}}\ \biggl(\begin{matrix}
\phantom{-}\sfrac1\vel{+}\ep^{2\h_2} & \im\sfrac{1-\vel}{\vel}\ep^{\h_2}\\[4pt]
-\im\sfrac{1-\vel}{\vel}\ep^{\h_2} & 1{+}\sfrac1\vel\ep^{2\h_2} 
\end{matrix}\biggr) \biggl(\begin{matrix}
1 & -\im\ep^{\h_1} \\[4pt] \im\ep^{\h_1} & \ep^{2\h_1}
\end{matrix}\biggr) \quad, \\[8pt]
\bigl(\unity-\sfrac{1+\vel}{\vel}\,P_1  \bigr)\,T_2\,\Tdag_2 \=&\
\frac{1}{1+\ep^{2\h_1}}\ \biggl(\begin{matrix}
-\sfrac1\vel{+}\ep^{2\h_1} & \im\sfrac{1+\vel}{\vel}\ep^{\h_1} \\[4pt]
-\im\sfrac{1+\vel}{\vel}\ep^{\h_1} & 1{-}\sfrac1\vel\ep^{2\h_1}
\end{matrix}\biggr) \biggl(\begin{matrix}
1 & \im\ep^{\h_2} \\[4pt] -\im\ep^{\h_2} & \ep^{2\h_2}
\end{matrix}\biggr) \quad,
\end{aligned}
\end{equation}
which combine to
\begin{align}
G &\= \frac{1}{\vel^{-2}(1{-}\ep^{\h_1+\h_2})^2+(\ep^{\h_1}{+}\ep^{\h_2})^2}\
\Biggl(\begin{matrix}
\vel^{-2}(1{-}\ep^{\h_1+\h_2})^2{-}(\ep^{\h_1}{+}\ep^{\h_2})^2 & 
2\im\vel^{-1}(\ep^{\h_1}{+}\ep^{\h_2})(1{-}\ep^{\h_1+\h_2}) \\[8pt]
2\im\vel^{-1}(\ep^{\h_1}{+}\ep^{\h_2})(1{-}\ep^{\h_1+\h_2}) &
\vel^{-2}(1{-}\ep^{\h_1+\h_2})^2{-}(\ep^{\h_1}{+}\ep^{\h_2})^2 
\end{matrix}\Biggr) \nonumber \\[12pt] &\=
\frac{1}{\sinh^2\g\vel t +\vel^2\cosh^2\g x}\ 
\Biggl(\begin{matrix}
\sinh^2\g\vel t-\vel^2\cosh^2\g x & 2\im\vel\sinh\g\vel t\ \cosh\g x\\[8pt]
2\im\vel\sinh\g\vel t\ \cosh\g x & \sinh^2\g\vel t-\vel^2\cosh^2\g
\end{matrix}\Biggr) \label{Gcomm}
\end{align}
with the help of
\begin{equation}
\h_1+\h_2 \= -2\g\vel t \und \h_1-\h_2 \= 2\g x \quad.
\end{equation}

Comparing with (\ref{embed}) we learn that, with $\p_+{-}\p_-=:4\b$,
\begin{equation}
\ep^{\im\b} \cos\sfrac{\vp}{2} \= 
\frac{\sinh^2\g\vel t-\vel^2\cosh^2\g x}{\sinh^2\g\vel t +\vel^2\cosh^2\g x}
\und
\ep^{\im\b} \sin\sfrac{\vp}{2} \=
\frac{2\vel\,\sinh\g\vel t\ \cosh\g x}{\sinh^2\g\vel t+\vel^2\cosh^2\g x}\quad,
\end{equation}
so that
\begin{equation}
\tan\sfrac{\vp}{4}\ \equiv\ \frac{\sin\sfrac{\vp}{2}}{1{+}\cos\sfrac{\vp}{2}}\=
\frac{2\vel\,\sinh\g\vel t\ \cosh\g x}
{(\ep^{\im\b}{+}1)\sinh^2\g\vel t + (\ep^{\im\b}{-}1)\,\vel^2\cosh^2\g x}
\end{equation}
analytically continues via $\vel{\to}\im\vel$ to
\begin{equation}
\tan\sfrac{\bar\vp}{4} \= \frac{2\vel\,\sin\gb\vel t\ \cosh\gb x}
{(\ep^{\im\b}{+}1)\sin^2\gb\vel t + (\ep^{\im\b}{-}1)\,\vel^2\cosh^2\gb x}
\end{equation}
with $\gb=\sfrac{1}{\sqrt{1+\vel^2}}$.
Since $\bar\vp$ is real we must have $\b{=}0$ or $\b{=}\pi$.
The boundary condition $\bar\vp\to0$ for $|x|\to\infty$ 
selects the second option,\footnote{
Alternatively, begin with $\p_+{=}\p_-{=}\vp$ and shift
$\p_\pm\to\p_\pm\pm2\pi$, or else, put $\b{=}0$ and shift $\vp\to\vp+2\pi$.}
and we have recovered the celebrated breather configuration~\cite{raja}
\begin{equation}
-\bar\vp \= 4\,\arctan\,\Bigl\{\frac{\sin\gb\vel t}{\vel\,\cosh\gb x}\Bigr\}
\quad.
\end{equation}

\bigskip

\section{Noncommutative construction}

\noindent
When attempting to repeat the above computation in the Moyal-deformed case,
one must account for the noncommutativity of the co-moving coordinates,
\begin{equation}
[t\,,\,x]_\star \= \im\theta \qquad\Longrightarrow\qquad
[\h_1\,,\,\h_2]_\star \= 2\im\theta\,\g^2\vel
\= 2\im\theta\,\sfrac{\vel}{1-\vel^2} \ =:\ \im\la \quad,
\end{equation}
which leads to the fundamental intertwining relation,
\begin{equation}
\ep^{(a_1+b_1)\h_1+(a_2+b_2)\h_2} \=
\ep^{-\frac{\im}{2}\la a\wedge b}\ 
\ep^{a_1\h_1+a_2\h_2}\star\ep^{b_1\h_1+b_2\h_2} \=
\ep^{\frac12(a_1\h_1+a_2\h_2)}\star
\ep^{b_1\h_1+b_2\h_2}\star
\ep^{\frac12(a_1\h_1+a_2\h_2)}
\end{equation}
which (for $f$ regular at zero) implies
\begin{equation}
\ep^{a_1\h_1+a_2\h_2} \star f(\ep^{b_1\h_1+b_2\h_2}) \=
f(\ep^{b_1\h_1+b_2\h_2+\im\la a\wedge b}) \star \ep^{a_1\h_1+a_2\h_2}
\quad.
\end{equation}

Again, we put $2\a=1$ for convenience.
The projectors~(\ref{projconf}) are unaffected by the deformation,
but the star products become relevant when $T_1$ or $P_1$ meets $T_2$ or~$P_2$.
As a basic ingredient in~(\ref{Gconf}), we first compute
\begin{align} \nonumber
\Tdag_j\star(\unity{-}\s P_k)\star T_j \=&\
\bigl( 1\,,\,-\im\ep^{\h_j} \bigr) \star \biggl[
(1{+}\ep^{2\h_k})^{-\frac12} \biggl( \begin{matrix}
1{-}\s{+}\ep^{2\h_k}\!\! & -\im\s\,\ep^{\h_k} \\[8pt] 
\!\!\im\s\,\ep^{\h_k} & 1{+}(1{-}\s)\ep^{2\h_k} \end{matrix} \biggr)\,
(1{+}\ep^{2\h_k})^{-\frac12} \biggr] \star \biggl( \begin{matrix}
1 \\[8pt] \im\ep^{\h_j} \end{matrix} \biggr) \\[8pt] \nonumber \=&\quad\ \;
(1{+}\ep^{2\h_k})^{-\frac12} \star 
(1{-}\s+\ep^{2\h_k}) \star
(1{+}\ep^{2\h_k})^{-\frac12} \\ \nonumber & +\
(1{+}\ep^{2\h_k})^{-\frac12} \star 
\s\,\ep^{\h_j+\h_k\mp\frac\im2\la} \star
(1{+}\ep^{2\h_k\mp2\im\la})^{-\frac12} \\ & +\
(1{+}\ep^{2\h_k\pm2\im\la})^{-\frac12} \star 
\s\,\ep^{\h_j+\h_k\pm\frac\im2\la} \star 
(1{+}\ep^{2\h_k})^{-\frac12} \\ \nonumber & +\
(1{+}\ep^{2\h_k\pm2\im\la})^{-\frac12} \star
(\ep^{\h_j}+(1{-}\s)\ep^{2\h_j+2\h_k}) \star
(1{+}\ep^{2\h_k\mp2\im\la})^{-\frac12}  \\[8pt] \nonumber \=&\
(1{+}\ep^{2\h_k})^{-\frac12} \star 
\bigl[(1{-}\s)(1{-}\ep^{\h_1+\h_2})^2 + (\ep^{\h_1}{+}\ep^{\h_2})^2\bigr]\star
(1{+}\ep^{2\h_k})^{-\frac12} \ +\ O(\la^2) \quad.
\end{align}
In the last step, we dropped terms of $O(\la^2)$ in order to arrive at a 
manageable expression. Inserting the above into (\ref{Gconf}) and abbreviating
\begin{equation} \label{abbr1}
N_k \= \Tdag_k T_k \= 1+\ep^{2\h_k} \und 
D   \= (1{-}\s)(1{-}\ep^{\h_1+\h_2})^2 + (\ep^{\h_1}{+}\ep^{\h_2})^2 \quad,
\end{equation}
we find the matrix elements of $G$ up to $O(\la^2)$ (denoted by `$\simeq$'),
\begin{align} &
\begin{aligned} \label{G11}
G_{11} \ \simeq\ &\ 1\ -\ 2\,N_2^{-1}\star \bigl[ 
\ep^{2\h_2}+\sfrac1\vel-\sfrac{1-\vel}{\vel}\ep^{\h_1+\h_2-\frac\im2\la}
\bigr]\star N_2^{\frac12}\star D^{-1}\star N_2^{\frac12} \\ 
& \quad\, -\ 2\,N_1^{-1}\star \bigl[ 
\ep^{2\h_1}-\sfrac1\vel+\sfrac{1+\vel}{\vel}\ep^{\h_1+\h_2+\frac\im2\la}
\bigr]\star N_1^{\frac12}\star D^{-1}\star N_1^{\frac12}
\quad,
\end{aligned} \\[8pt] &
\begin{aligned}
G_{12} \ \simeq\ &\qquad\phantom{(} 2\im\,N_2^{-1}\star \bigl[ 
\ep^{2\h_2}+\sfrac1\vel-\sfrac{1-\vel}{\vel}\ep^{\h_1+\h_2-\frac\im2\la}
\bigr]\star N_2^{\frac12}\star D^{-1}\star N_2^{\frac12}\star\ep^{\h_1}\\
& \quad\, - 2\im\,N_1^{-1}\star \bigl[ 
\ep^{2\h_1}-\sfrac1\vel+\sfrac{1+\vel}{\vel}\ep^{\h_1+\h_2+\frac\im2\la}
\bigr]\star N_1^{\frac12}\star D^{-1}\star N_1^{\frac12}\star\ep^{\h_2}
\quad,
\end{aligned} \\[8pt] &
\begin{aligned}
G_{21} \ \simeq\ & \quad\, -2\im\,N_2^{-1}\star \bigl[ 
\ep^{\h_1}-\sfrac{1-\vel}{\vel}\ep^{\h_2}+\sfrac1\vel\ep^{\h_1+2\h_2-\im\la}
\bigr]\star N_2^{\frac12}\star D^{-1}\star N_2^{\frac12}\\
& \quad\, + 2\im\,N_1^{-1}\star \bigl[ 
\ep^{\h_2}+\sfrac{1+\vel}{\vel}\ep^{\h_1}-\sfrac1\vel\ep^{2\h_1+\h_2+\im\la}
\bigr]\star N_1^{\frac12}\star D^{-1}\star N_1^{\frac12}
\quad,
\end{aligned} \\[8pt] &
\begin{aligned} \label{G22}
G_{22} \ \simeq\ &\ 1\ -\ 2\,N_2^{-1}\star \bigl[ 
\ep^{\h_1}-\sfrac{1-\vel}{\vel}\ep^{\h_2}+\sfrac1\vel\ep^{\h_1+2\h_2-\im\la}
\bigr]\star N_2^{\frac12}\star D^{-1}\star N_2^{\frac12}\star\ep^{\h_1}\\
& \quad\, -\ 2\,N_1^{-1}\star \bigl[ 
\ep^{\h_2}+\sfrac{1+\vel}{\vel}\ep^{\h_1}-\sfrac1\vel\ep^{2\h_1+\h_2+\im\la}
\bigr]\star N_1^{\frac12}\star D^{-1}\star N_1^{\frac12}\star\ep^{\h_2}
\quad.
\end{aligned} 
\end{align}
There is some pattern with respect to the interchange 
$\h_1\leftrightarrow\h_2$ and regarding sign flips of $\vel$ and~$\la$,
but no obvious symmetry under $\th\to-\th$. We have chosen the positions
of the $N_k$ such that their arguments are not shifted. The equalities
$G_{11}=G_{22}$ and $G_{12}=G_{21}$ are far from manifest. Note that, for the
exact result, $D$ is to be inverted with respect to star multiplication.
However, since $D_\star^{-1}-D^{-1}=O(\la^2)$, we may take the ordinary inverse
in (\ref{G11})--(\ref{G22}).
Finally, the commutative limit collapses $G$ to~(\ref{Gcomm}), 
since all $N_k$~factors cancel and disappear.

We can also employ the last equation of (\ref{Gconf}), which expresses~$G$
in terms of projectors only. After rescaling the projectors to
\begin{equation}
\tP_1 \=  \sfrac{1-\vel}{\vel}\,P_1 \und
\tP_2 \= -\sfrac{1+\vel}{\vel}\,P_2 \qquad\textrm{such that}\qquad
\tP_1\star\tP_2 \= \s\,P_1\star P_2 \quad,
\end{equation}
we rewrite
\begin{equation}
\begin{aligned}
G \=& \unity\ -\ 
2\,P_1\star\bigl[\unity-\tP_2\star\tP_1\bigr]_\star^{-1}
\star\bigl[1+\tP_2\bigr]\ -\
2\,P_2\star\bigl[\unity-\tP_1\star\tP_2\bigr]_\star^{-1}
\star\bigl[1+\tP_1\bigr] \\[4pt] \=& \unity\ -\ 
\sfrac{2\vel}{1-\vel}\,\bigl[ 
\tP_1 + \tP_1{\star}\tP_2 + \tP_1{\star}\tP_2{\star}\tP_1 + \ldots \bigr] \ +\
\sfrac{2\vel}{1+\vel}\,\bigl[ 
\tP_2 + \tP_2{\star}\tP_1 + \tP_2{\star}\tP_1{\star}\tP_2 + \ldots \bigr]
\quad.
\end{aligned}
\end{equation}
In the last line, 
we have traded the notorious star inverses for formal geometric series,
\begin{equation}
\bigl[ \unity - \tP_j\star\tP_k\bigr]_\star^{-1} \=
\sum_{n=0}^\infty\,\bigl[ \tP_j\star\tP_k\bigr]^n_\star 
\qquad\textrm{with}\quad (j,k) = (1,2)\ \textrm{or}\ (2,1) \quad,
\end{equation}
which may be truncated in an approximation for large velocities $\vel\to1$. 
In this way, $G$ is given as a power series in words 
$P_j\star P_k\star P_j\star\ldots\star P_\ell$.
Remembering~(\ref{proj}) and abbreviating also
\begin{equation} \label{abbr2}
N_{jk} \= \Tdag_j\star T_k \= 1 -\ \ep^{\h_j}\star\ep^{\h_k} 
\= 1 -\ \ep^{\h_1+\h_2\pm\frac\im2\la} \quad,
\end{equation}
the `projector words' simplify to
\begin{equation}
\begin{aligned}
P_j\star P_k\star P_j\star\cdots\star P_\ell \=
\Bigl( \begin{matrix} 1 \\ \pm\im\ep^{\h_j} \end{matrix} \Bigr)\,
N_j^{-1}\star N_{jk}\star N_k^{-1}\star N_{kj}\star\cdots\star N_\ell^{-1}\, 
\bigl( 1 \quad {\pm}\im\ep^{\h_\ell} \bigr) \phantom{\quad,} \\[4pt] 
\= \Bigl( \begin{matrix} \ep^{-\frac12\h_j} \\ \pm\im\ep^{\frac12\h_j}
\end{matrix} \Bigr)\,
\tN_j^{-1}\star\tN_{jk}\star\tN_k^{-1}\star\tN_{kj}
\star\cdots\star\tN_\ell^{-1}\,
\bigl( \ep^{-\frac12\h_\ell} \quad {\pm}\im\ep^{\frac12\h_\ell} \bigr) \quad,
\end{aligned}
\end{equation}
where the last line is a symmetric rewriting with
\begin{equation}
\tN_k \= \ep^{-\h_k}+\ep^{\h_k} \und
\tN_{jk} \= \ep^{-\frac12\h_j}\star\ep^{-\frac12\h_k}
- \ep^{\frac12\h_j}\star\ep^{\frac12\h_k} 
\= \ep^{\pm\frac\im8\la}
\bigl( \ep^{-\frac12(\h_1+\h_2)}-\ep^{\frac12(\h_1+\h_2)} \bigr) \quad.
\end{equation}
Pulling all together, one arrives at
\begin{align} \label{Gseries}
G_{11} =&\ 1 
-2N_1^{-1} 
-2N_2^{-1}
+2\sfrac{1+\vel}{\vel}N_1^{-1}{\star}N_{12}{\star}N_2^{-1}
-2\sfrac{1-\vel}{\vel}N_2^{-1}{\star}N_{21}{\star}N_1^{-1}
+\ldots \quad, \\[8pt] \nonumber
G_{12} =& \quad\phantom{xr}
 2\im N_1^{-1}\ep^{\h_1}
-2\im N_2^{-1}\ep^{\h_2} 
+2\im\sfrac{1+\vel}{\vel}N_1^{-1}{\star}N_{12}{\star}N_2^{-1}\ep^{\h_2}
+2\im\sfrac{1-\vel}{\vel}N_2^{-1}{\star}N_{21}{\star}N_1^{-1}\ep^{\h_1}
+\ldots \quad, \\[8pt] \nonumber
G_{21} =& \quad\! 
-2\im\ep^{\h_1}N_1^{-1} 
+2\im\ep^{\h_2}N_2^{-1}
+2\im\sfrac{1+\vel}{\vel}\ep^{\h_1}N_1^{-1}{\star}N_{12}{\star}N_2^{-1}
+2\im\sfrac{1-\vel}{\vel}\ep^{\h_2}N_2^{-1}{\star}N_{21}{\star}N_1^{-1}
+\ldots \quad, \\[8pt] \nonumber
G_{22} =&\ 1 
-2\ep^{\h_1}N_1^{-1}\ep^{\h_1}
-2\ep^{\h_2}N_2^{-1}\ep^{\h_2}
-2\sfrac{1+\vel}{\vel}\ep^{\h_1}N_1^{-1}{\star}N_{12}{\star}N_2^{-1}\ep^{\h_2}
+2\sfrac{1-\vel}{\vel}\ep^{\h_2}N_2^{-1}{\star}N_{21}{\star}N_1^{-1}\ep^{\h_1}
+\!\ldots
\end{align}
with $N_k$ and $N_{jk}$ to be taken from (\ref{abbr1}) and (\ref{abbr2}),
respectively. This is an exact result. No star-inverse needs to be taken,
but we are left with infinite series, which may be summed in closed form
only for $\theta{=}0$.

\bigskip

\section{Expanding in $\theta$}

\noindent
The task is to extract the deformed breather configuration
\begin{equation}
\ep_\star^{\pm\frac\im2\phi_\pm} \= g_\pm 
\= G_{11} \pm G_{21} \= G_{11} \pm G_{12}\ =:\ G_e \pm G_o
\end{equation}
from (\ref{G11})--(\ref{G22}) or from (\ref{Gseries}), 
at least to subleading order in a $\theta$~expansion,
\begin{equation}
f \= f^{(0)} + \la\,f^{(1)} + \la^2 f^{(2)} + \ldots 
\ \simeq\ f^{(0)} + \la\,f^{(1)}
\qquad\textrm{for}\qquad f\in\{G,g_\pm,\phi_\pm,\ldots\}\quad.
\end{equation}
Keeping $\vel$ fixed and noticing that 
$\ep_\star^h = \ep_{\phantom{\star}}^h + O(\la^2)$ for any function~$h$, 
we have
\begin{equation}
\p_\pm \ \simeq\ \mp 2\im\ln g_\pm \ \simeq\ 
\mp 2\im\ln (g_\pm^{(0)}\!+\la g_\pm^{(1)}) \ \simeq\
\mp 2\im\ln g_\pm^{(0)} \mp 2\im\la\,g_\pm^{(1)}\!/g_\pm^{(0)} \=
\p_\pm^{(0)} \mp 2\im\la\,g_\pm^{(1)} \ep^{\mp\frac\im2\phi_\pm^{(0)}}
\end{equation}
and thus
\begin{equation} \label{anglexp}
\begin{aligned}
\sfrac12(\p_+{+}\p_-) \ &\simeq\ \vp\ +\ \,
\im\la\,g_+^{(1)}\ep^{-\frac\im2\vp}\ -\ 
\im\la\,g_-^{(1)}\ep^{\frac\im2\vp} \= 
\vp\ +\ \,2\la\,G_e^{(1)}\sin\sfrac{\vp}{2}
\ +\ 2\im\la\,G_o^{(1)}\cos\sfrac{\vp}{2} \quad, \\[4pt]
\sfrac12(\p_+{-}\p_-) \ &\simeq\ 2\pi +\
\im\la\,g_+^{(1)}\ep^{-\sfrac\im2\vp}\ +\ 
\im\la\,g_-^{(1)}\ep^{\frac\im2\vp} \=
2\pi +\ 2\im\la\,G_e^{(1)}\cos\sfrac{\vp}{2}
\ +\ 2\la\,G_o^{(1)}\sin\sfrac{\vp}{2}
\end{aligned}
\end{equation}
since $\phi_\pm^{(0)}=\vp\pm 2\pi$. 
{}From (\ref{Gseries}) one learns that, in the $\la$~expansion,
the even orders of $G_e$ and the odd orders of $G_o$ are real while the
odd orders of $G_e$ and the even orders of $G_o$ are imaginary.
Because (\ref{anglexp}) must be real equations for $G\in\tU(1){\times}\tU(1)$,
this implies that
\begin{equation} \label{eorel}
G_e^{(1)}\sin\sfrac{\vp}{2}\ +\ \im\,G_o^{(1)}\cos\sfrac{\vp}{2} \=0
\qquad\Longrightarrow\qquad
\p_+^{(1)} = -\p_-^{(1)} 
= 2\im\,G_e^{(1)}/\cos\sfrac{\vp}{2} 
= 2\,G_o^{(1)}/\sin\sfrac{\vp}{2}\quad,
\end{equation}
and so the sine-Gordon field $\sfrac12(\p_+{+}\p_-)$ gets deformed only
at $O(\la^2)$ while the orthogonal combination $\sfrac12(\p_+{-}\p_-)$
is turned on at~$O(\la)$. Interestingly, the relation~(\ref{eorel}) is 
again the commutative one, thus
\begin{equation}
G\ \equiv\ \biggl(\begin{matrix}G_e & G_o \\[4pt] G_o & G_e\end{matrix}\biggr)
\ \simeq\ \ep^{\im\pi+\im\la\chi}\ 
\biggl(\begin{matrix}
\  \cos\sfrac{\vp}{2} & \im\sin\sfrac{\vp}{2} \\[4pt] 
\im\sin\sfrac{\vp}{2} & \  \cos\sfrac{\vp}{2}
\end{matrix}\biggr) \qquad\textrm{with}\quad 
\chi = \im\,G_e^{(1)}/\cos\sfrac{\vp}{2} = G_o^{(1)}/\sin\sfrac{\vp}{2} \quad.
\end{equation}
For computing $\chi$ it suffices to look at any one of the $G$~matrix elements.

In order to expand $G$ to $O(\la)$ we need the first subleading term in
multiple star products,
\begin{equation}
f_1\star f_2\star f_3\star\cdots\star f_n \ \simeq\ 
f_1\ f_2\ f_3 \cdots f_n \ +\ \sfrac\im2 \la \sum_{i<j} 
f_1 \cdots (\pa_{[1}f_i) \cdots (\pa_{2]} f_j) \cdots f_n \quad,
\end{equation}
where \ 
$(\pa_{[1}f_i)(\pa_{2]}f_j)\equiv\frac{\pa f_i}{\pa\h_1}\frac{\pa f_j}{\pa\h_2}
-\frac{\pa f_i}{\pa\h_2}\frac{\pa f_j}{\pa\h_1}$.
The products appearing in (\ref{G11})--(\ref{G22}) take the forms
\begin{align}
N_2^{-1}{\star}\ep^h{\star}N_2^{\frac12}{\star}D^{-1}{\star}N_2^{\frac12} 
&\ \simeq\ \ep^h D^{-1}\ +\ \sfrac\im2\la\,\ep^h D^{-2} \bigl(
2D\,\pa_1 h\,\sfrac{N_2'}{N_2}-\pa_1 D\,\sfrac{N_2'}{N_2}-\pa_{[1}h\,\pa_{2]}D
\bigr) \quad,\\[4pt]
N_1^{-1}{\star}\ep^h{\star}N_1^{\frac12}{\star}D^{-1}{\star}N_1^{\frac12} 
&\ \simeq\ \ep^h D^{-1}\ -\ \sfrac\im2\la\,\ep^h D^{-2} \bigl(
2D\,\pa_2 h\,\sfrac{N_1'}{N_1}-\pa_2 D\,\sfrac{N_1'}{N_1}-\pa_{[2}h\,\pa_{1]} D
\bigr) \quad,\\[4pt]
N_2^{-1}{\star}\ep^h{\star}N_2^{\frac12}{\star}D^{-1}{\star}N_2^{\frac12}
{\star}\ep^{\h_1} &\ \simeq\ \ep^{h+\h_1}D^{-1} \\
&\ +\ \sfrac\im2\la\,\ep^{h+\h_1} D^{-2} \bigl(
2D\,\pa_1 h\,\sfrac{N_2'}{N_2}-\pa_1 D\,\sfrac{N_2'}{N_2}-\pa_{[1}h\,\pa_{2]}D
+\pa_2 D-D\,\pa_2 h \bigr) \quad, \nonumber\\[4pt]
N_1^{-1}{\star}\ep^h{\star}N_1^{\frac12}{\star}D^{-1}{\star}N_1^{\frac12}
{\star}\ep^{\h_2} &\ \simeq\ \ep^{h+\h_2}D^{-1} \\
&\ -\ \sfrac\im2\la\,\ep^{h+\h_2} D^{-2} \bigl(
2D\,\pa_2 h\,\sfrac{N_1'}{N_1}-\pa_2 D\,\sfrac{N_1'}{N_1}-\pa_{[2}h\,\pa_{1]}D
+\pa_1 D-D\,\pa_1 h \bigr) \quad, \nonumber
\end{align}
with $h$ being linear in $\h_1$ and $\h_2$. Collecting all terms and
noticing cancellations we obtain
\begin{equation}
\begin{aligned}
\im\vel D^2 G_{11}^{(1)} &\= 
 - \pa_1D\sfrac{N_2'}{N_2}
 - \ep^{\h_1+\h_2}\bigl( 
 \pa_1D(1{-}\sfrac{N_2'}{N_2}) - \pa_2D - D(1{-}2\sfrac{N_2'}{N_2})\bigr)
 \ +\ (1\leftrightarrow2) \quad, \\[4pt]
-D^2 G_{12}^{(1)} &\=
\ep^{\h_1+2\h_2}\bigl(\pa_1D(2{-}\sfrac{N_2'}{N_2})+\pa_2D-2D\bigr) 
 + \ep^{2\h_1+\h_2}(\pa_1D{-}2D)(1{-}\sfrac{N_2'}{N_2})
 \ +\ (1\leftrightarrow2) \quad, \\[4pt]
 D^2 G_{21}^{(1)} &\=
\ep^{\h_1}\bigl((2D{-}\pa_1D)\sfrac{N_2'}{N_2}-\pa_2D\bigr)
 + \ep^{\h_2}\pa_1D(1{-}\sfrac{N_2'}{N_2})
 \ +\ (1\leftrightarrow2) \quad, \\[4pt]
\im\vel D^2 G_{22}^{(1)} &\=
\ep^{2\h_1+2\h_2}(\pa_1D{-}2D)(2{-}\sfrac{N_2'}{N_2})
 - \ep^{\h_1+\h_2}\bigl( 
 \pa_1D(1{-}\sfrac{N_2'}{N_2})+\pa_2D-D\bigr)
 \ +\ (1\leftrightarrow2) \quad, \\[4pt]
\end{aligned}
\end{equation}
\vskip-12pt\noindent
which further collapses to
\begin{equation}
\begin{aligned}
G_e^{(1)} &\=
-2\im\vel\,\ep^{\h_1+\h_2} 
\frac{(1{-}\ep^{\h_1+\h_2})^2-\vel^2(\ep^{\h_1}+\ep^{\h_2})}
     {[(1{-}\ep^{\h_1+\h_2})^2+\vel^2(\ep^{\h_1}+\ep^{\h_2})]^2} \=
-\frac{\im\vel}{2}
\frac{\sinh^2\g\vel t-\vel^2\cosh^2\g x}{[\sinh^2\g\vel t+\vel^2\cosh^2\g x]^2}
\quad, \\[8pt]
G_o^{(1)} &\=
\ \ \,4\vel^2\ep^{\h_1+\h_2} 
\frac{(\ep^{\h_1}+\ep^{\h_2})(1{-}\ep^{\h_1+\h_2})}
     {[(1{-}\ep^{\h_1+\h_2})^2+\vel^2(\ep^{\h_1}+\ep^{\h_2})]^2} \,\=
\ \vel^2 \frac{\sinh\g\vel t\ \cosh\g x}{[\sinh^2\g\vel t+\vel^2\cosh^2\g x]^2}
\quad.
\end{aligned}
\end{equation}
Comparing to (\ref{Gcomm}), we indeed confirm that \ 
$G^{(1)}=\im\,\chi\,G^{(0)}$, and hence
\begin{equation}
g_{\pm} \ \simeq\ \ep^{\im\pi+\im\la\chi}\ \ep^{\pm\frac\im2\vp}
\qquad\textrm{with}\qquad
\chi \= \frac{-\vel/2}{\sinh^2\g\vel t+\vel^2\cosh^2\g x} \quad.
\end{equation}
It appears as if the sine-Gordon field gets deformed via \
$\vp\to\vp\mp2\la\chi$, but this is misleading.

This formula provides the explicit $O(\theta)$ correction to the
commutative kink-antikink configuration. To obtain the breather,
we still must analytically continue~$\vel\to\im\vel$, which yields
\begin{equation}
\lambda\ \to\ 2\,\theta\sfrac{\im\,\vel}{1+\vel^2} \ =:\ \im\,\bar\lambda
\und \chi\ \to\ \im\,\frac{\vel/2}{\sin^2\gb\vel t+\vel^2\cosh^2\gb x}
\ =:\ \im\,\bar\chi \quad,
\end{equation}
so that the leading correction to~$G$ remains a phase factor.
This is the main result of this letter. Clearly, $\chi$ oscillates
with twice the classical breather frequency $\omega=\bar\gamma\,\vel$.
More generally, our construction shows that the deformed breather frequency
does not depend on~$\theta$ at all. Below we illustrate the shapes of
$\bar\vp(t,x)$ and $\bar\chi(t,x)$ for a typical value of~$\vel$.

\bigskip

\noindent
{\bf Acknowledgements}

\noindent
The authors are grateful to D.~Harland, S.~K\"{u}rk\c{c}\"{u}o\v{g}lu 
and A.D.~Popov for fruitful discussions.
The work of A.F. and O.L.~is partially supported by the Deutsche
Forschungsgemeinschaft.
\bigskip
%\newpage

%
\begin{figure}[ht]
\psfrag{t}{${}\ \,\textstyle{t}$}
\psfrag{x}{\lower10pt\hbox{\hspace{4pt}$\textstyle{x}$}}
\psfrag{phi}{$\hspace{-4pt}\textstyle{\bar\vp}$}
\centerline{\includegraphics[width=8cm]{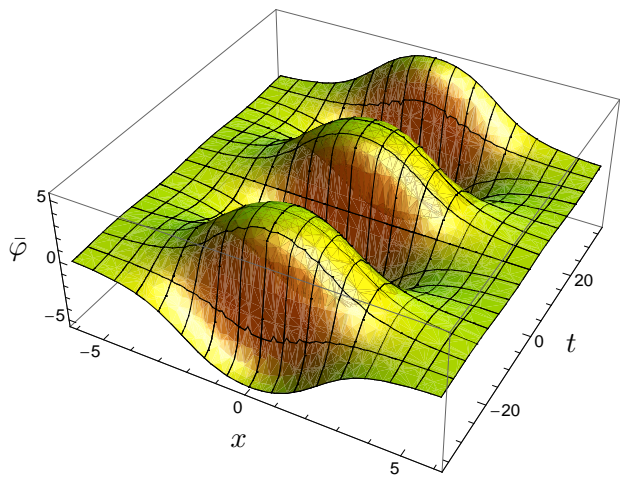}}
\caption{Commutative breather $\bar\vp(t,x)$ for $\vel=0.21$}
\label{fig:1}
\end{figure}
\begin{figure}[ht]
\psfrag{t}{${}\ \ \textstyle{t}$}
\psfrag{x}{\lower11pt\hbox{\hspace{2pt}$\textstyle{x}$}}
\psfrag{chi}{$\hspace{-5pt}\textstyle{\bar\chi}$}
\centerline{\includegraphics[width=8cm]{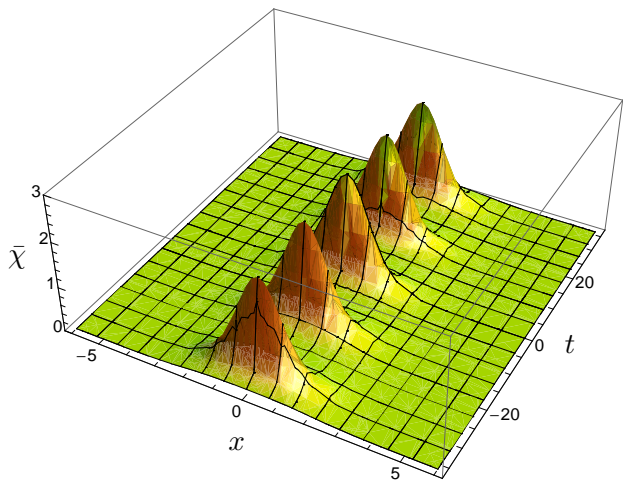}}
\caption{Noncommutative correction $\bar\chi(t,x)$ for $\vel=0.21$}
\label{fig:2}
\end{figure}

\end{document}